\documentclass[a4paper]{scrartcl}
\usepackage{scrpage2}
\usepackage{moreverb}
\usepackage{amsmath,amssymb,graphicx}
\usepackage{hyperref}
\DeclareMathAlphabet\mathbfcal{OMS}{cmsy}{b}{n}

\newcommand{\affil}[1]{$^{#1}$}
\newcommand{\affilnum}[1]{#1~}
\newcommand{\corrauth}{$^{*}$}

\begin{document}
\pagestyle{scrheadings}
\chead{\footnotesize J.~Dutin\'e, et al. -- Explicit Time Integration of Transient Eddy Current Problems}

\title{Explicit Time Integration of Transient Eddy Current Problems}

\author{J.~Dutin\'e\affil{1}\corrauth, M.~Clemens\affil{1}, S.~Sch{\"o}ps\affil{2}, G.~Wimmer\affil{3}}

\date{\small%
\affilnum{1}Chair of Electromagnetic Theory, School of Electrical, Information and Media Engineering, University of Wuppertal, Rainer-Gruenter-Str. 21, 42119 Wuppertal, Germany\break \affilnum{2}Graduate School CE, Technische Universit{\"a}t Darmstadt, Dolivostr. 15, 64293 Darmstadt, Germany\break\affilnum{3}Fakult{\"a}t f{\"u}r angewandte Natur- und Geisteswissenschaften, Hochschule f{\"u}r angewandte Wissenschaften W{\"u}rzburg-Schweinfurt, Ignaz-Sch{\"o}nstr. 11, 97421 Schweinfurt, Germany\\[2em]
}

\publishers{\parbox{0.8\linewidth}
{\small\textbf{Abstract. }
For time integration of transient eddy current problems commonly implicit time integration methods are used, where in every time step one or several nonlinear systems of equations have to be linearized with the Newton-Raphson method due to ferromagnetic materials involved. In this paper, a generalized Schur-complement is applied to the magnetic vector potential formulation, which converts a differential-algebraic equation system of index 1 into a system of ordinary differential equations (ODE) with reduced stiffness. For the time integration of this ODE system of equations, the explicit Euler method is applied. The Courant-Friedrich-Levy (CFL) stability criterion of explicit time integration methods may result in  small time steps. Applying a pseudo-inverse of the discrete curl-curl operator in nonconducting regions of the problem is required in every time step. For the computation of the pseudo-inverse, the preconditioned conjugate gradient (PCG) method is used. The cascaded Subspace Extrapolation method (CSPE) is presented to produce suitable start vectors for these PCG iterations. The resulting scheme is validated using the TEAM 10 benchmark problem.}}

\maketitle

\let\thefootnote\relax\footnotetext{This is the pre-peer reviewed version of the article, which has been accepted for publication by the International Journal of Numerical Modelling: Electronic Networks, Devices and Fields (ISSN: 1099-1204). This article may be used for non-commercial purposes in accordance with Wiley Terms and Conditions for Self-Archiving.}

\section{Introduction}

In the computation of magnetoquasistatic fields spatial discretization e.g by  the Finite Element Method (FEM) of the magnetic vector potential formulation yields a differential-algebraic equation system of index 1 (DAE(1)) \cite{Schur2}. This DAE(1) system is of infinite stiffness and can only be integrated in time by suitable, unconditionally stable implicit time integration methods.
Frequently used implicit time integration methods are e.g. the implicit Euler method or the singly diagonal implicit Runge-Kutta schemes (SDIRK) \cite{HairerWanner}.
The nonlinear BH-characteristic in ferromagnetic materials requires the solution of at least one large nonlinear equation system in every time step. For this, the Newton-Raphson method is an established linearization scheme which may require several iterations per time step. In each Newton-Raphson iteration the stiffness-matrix and the Jacobian-matrix need to be reassembled or at least updated and the resulting linear algebraic equation system has to be solved. As these systems may be high dimensional for 3D problems, each time step is rather time consuming and the possibilities for acceleration and parallelization involve complicated numerical linear algebra techniques.

The use of the Newton-Raphson method can be avoided within explicit time integration schemes which are not standard schemes for eddy current problems.
An explicit time integration approach has been first proposed in \cite{Yioultsis01}, where the explicit Finite Difference Time Domain (FDTD) method is used in the conductive regions of the problem. Here, in the nonconductive regions the boundary element method (BEM) is applied  for computing the air parts of the solution, corresponding to a magnetostatic Poisson problem \cite{Yioultsis01}. As an alternative to the BEM, the use of an adapted Perfectly Matched Layer (PML) is proposed in \cite{Yioultsis02}. In \cite{Auserhofer1} and \cite{Auserhofer2}, the conductive and nonconductive regions are also treated differently. Here, within a Discontinuous Galerkin FEM-framework the explicit time integration method is used for regions with conductive materials, while an FEM formulation with continuous ansatz functions and an implicit time integration method are applied to the nonconductive regions \cite{Auserhofer1},\cite{Auserhofer2}.\\
The work presented in this paper is originally based on an approach proposed in \cite{Schur2} and \cite{Schur1} based on a Schur-complement reformulation of the magnetoquasistatic problem. In this paper, the use of a generalized Schur-complement is proposed in which a pseudo-inverse of the singular curl-curl matrix is considered in the nonconductive regions.
The solution of the resulting system of equations is obtained by the preconditioned conjugate gradient (PCG) method. As this results in a multiple right-hand side problem, an optimized starting vector for the PCG-method can be computed by the subspace projection extrapolation method (SPE) \cite{SPE04}.
The algorithm of the SPE has been  modified to a cascaded SPE (CSPE) scheme that is used throughout this work. The generalized Schur-complement formulation will be explained in Section \ref{formulation}, SPE and CSPE will be outlined in Section \ref{CSPE} and the corresponding numerical results will be presented in Section \ref{results}.

\section{Mathematical Formulation}
\label{formulation}

The partial differential equation describing maqnetoquasistatic problems using the magnetic vector potential is given by:

\begin{equation}
\label{mqspde}
\kappa\dfrac{\partial{\vec{\mathrm{A}}(\mathrm{t})}}{\partial \mathrm{t}}+\vec{\nabla}\times\left(\nu\left(\vec{\mathrm{A}}(\mathrm{t})\right)\vec{\nabla}\times\vec{\mathrm{A}}(\mathrm{t})\right)=\vec{\mathrm{J}}_{\mathrm{s}}(\mathrm{t}),
\end{equation} in which $\kappa$ is the electrical conductivity, $\vec{\mathrm{A}}(\mathrm{t})$ is the time-dependent magnetic vector potential, $\nu$ is the reluctivity which may be nonlinear in ferromagnetic materials and $\vec{\mathrm{J}}_{\mathrm{s}}(\mathrm{t})$ is the time-dependent source current density.

Spatial discretization of \eqref{mqspde} by e.g. FEM \cite{FEM} or the finite integration technique (FIT) \cite{FIT}, results in a differential-algebraic system of equations of index 1 (DAE(1)) of the form

\begin{equation}
\label{eqDAE}
\mathbf{M}\dfrac{\mathrm{d}}{\mathrm{d}t}\mathbf{a}+\mathbf{K}(\mathbf{a})\mathbf{a}=\mathbf{j}_{s},
\end{equation} where \textbf{M} is the mass-matrix of conductivities, \textbf{a} is the vector of the degrees of freedom (dofs) corresponding to the magnetic vector potential, \textbf{K} is the stiffness-matrix of reluctivities corresponding to the singular curl-curl operator discretization and $\mathbf{j}_{s}$ is the transient source current density. A solenoidal right-hand side in  \eqref{eqDAE} ensures the existence of a solution, which is achieved if the source current density equals the curl of the electric vector potential \cite{Ren}.
 
The dofs vector $\mathbf{a}$ in \eqref{eqDAE} is reordered into a part $\mathbf{a}_{n}$ corresponding to dofs in the nonconductive regions and into a part $\mathbf{a}_{c}$ corresponding to dofs in conductive regions. The matrices $\mathbf{K}$ and $\mathbf{M}$ are decomposed accordingly and yield the reformulation of \eqref{eqDAE} with

\begin{equation}
\label{matrixdecomp}
\begin{bmatrix}
\mathbf{M}_{cc} & 0 \\
0 & 0
\end{bmatrix} \dfrac{\mathrm{d}}{\mathrm{d}t} \begin{bmatrix}
\mathbf{a}_{c}\\
\mathbf{a}_{n}
\end{bmatrix} + \begin{bmatrix}
\mathbf{K}_{cc}(\mathbf{a}_{c}) & \mathbf{K}_{cn}\\
\mathbf{K}^{\mathrm{T}}_{cn} & \mathbf{K}_{nn}
\end{bmatrix} \begin{bmatrix}
\mathbf{a}_{c}\\
\mathbf{a}_{n}
\end{bmatrix} = \begin{bmatrix}
0\\
\mathbf{j}_{s,n} 
\end{bmatrix},
\end{equation} where $\mathbf{M}_{cc}$ is the conductivity matrix, $\mathbf{K}_{cc}$ is the part of the curl-curl matrix in conductive media,  $\mathbf{K}_{nn}$ is the singular curl-curl part in air and $\mathbf{K}_{cn}$ and is the coupling part of $\mathbf{K}$ between both conducting and non-conducting regions,  and $\mathbf{j}_{s,n}$ is  the source current density in the nonconductive region.
Applying the Schur-complement to \eqref{matrixdecomp}, transforms the infinitely stiff DAE(1) into an ordinary differential equation system (ODE) of finite stiffness \cite{Schur1} 

\begin{subequations}
\begin{eqnarray}
\mathbf{M}_{cc}\dfrac{\mathrm{d}}{\mathrm{d}t}\mathbf{a}_{c}+\left[\mathbf{K}_{cc}(\mathbf{a}_{c})-\left(\mathbf{K}_{cn}\mathbf{K}^{\#}_{nn}\mathbf{K}^{\mathrm{T}}_{cn}\right)\right]\mathbf{a}_{c} &=& -\mathbf{K}_{cn}\mathbf{K}^{\mathrm{-1}}_{nn}\mathbf{j}_{s,n}, \label{ODE_1}\\
\mathbf{a}_{n} &=& \mathbf{K}^{\#}_{nn}\mathbf{j}_{s,n}-\mathbf{K}^{\#}_{nn}\mathbf{K}^{\mathrm{T}}_{cn}\mathbf{a}_{c}, \label{ODE_2}
\end{eqnarray}
\end{subequations} where $\mathbf{K}^{\#}_{nn}$ is the matrix representation of the pseudo-inverse of $\mathbf{K}_{nn}$. The matrix product $\mathbf{K}_{cn}\mathbf{K}^{\#}_{nn}\mathbf{K}^{\mathrm{T}}_{cn}$ is the generalized Schur complement.

Equation \eqref{ODE_1} represents an ODE in which the explicit Euler method with time step width $\Delta t$ can be used. In the $(m+1)$-th time step 
\begin{equation}
\mathbf{a}^{m+1}_{c} \mathrel{\mathop:} = \mathbf{a}^{m}_{c}+\Delta t\mathbf{M}^{\mathrm{-1}}_{cc}\left[\mathbf{K}_{cn}\mathbf{K}^{\#}_{nn}\mathbf{j}^{m+1}_{s,n}-\left(\mathbf{K}_{cc}(\mathbf{a}^{m}_{c})-\mathbf{K}_{cn}\mathbf{K}^{\#}_{nn}\mathbf{K}^{\mathrm{T}}_{cn}\right)\mathbf{a}^{m}_{c}\right], \label{amp1}
\end{equation} is computed. With $\mathbf{a}^{m+1}_{c}$ the solution vector $\mathbf{a}^{m+1}_{n}$ for the degrees of freedom in the nonconductive region can be separately calculated with

\begin{equation}
\mathbf{a}^{m+1}_{n} \mathrel{\mathop:} = \mathbf{K}^{\#}_{nn}\mathbf{j}^{m+1}_{s,n}-\mathbf{K}^{\#}_{nn}\mathbf{K}^{\mathrm{T}}_{cn}\mathbf{a}^{m+1}_{c}. \label{amp2}
\end{equation}

The singular matrix $\mathbf{K}_{nn}$ could be regularized with a grad-div-regularization resulting in a discrete vector Laplacian operator in free space \cite{Schur1}. As this is computationally expensive \cite{Schur1}, here the computation of a pseudo-inverse is proposed using the PCG method. In \eqref{amp1} and \eqref{amp2}, the pseudo-inverse is evaluated by solving systems of the kind:
\begin{subequations}
\begin{eqnarray}
\label{pseudo}
\mathbf{K}_{nn}\mathbf{k}_{p} &=& \mathbf{r}, \label{pseudo1}\\
\mathbf{k}_{p} &\mathrel{\mathop:} =& \mathbf{K}^{\#}_{nn}\mathbf{r}, \label{pseudo2}
\end{eqnarray}
\end{subequations} where $\mathbf{r}$ represents one of the vectors with which $\mathbf{K}^{\#}_{nn}$ is multiplied in \eqref{amp1}, \eqref{amp2}, i.e. $\mathbf{j}^{m+1}_{s,n}$, $\mathbf{K}^{\mathrm{T}}_{cn}\mathbf{a}^{m}_{c}$ and  $\mathbf{K}^{\mathrm{T}}_{cn}\mathbf{a}^{m+1}_{c}$. The matrix $\mathbf{K}^{\#}_{nn}$  does not need to be computed explicitely. Alternatively, a vector $\mathbf{k}_{p}$ according to \eqref{pseudo1}, \eqref{pseudo2} is computed by the PCG method for each right-hand side $\mathbf{r}$ stated above replacing a matrix vector multiplication $\mathbf{K}^{\#}_{nn}\mathbf{r}$ in \eqref{amp1}, \eqref{amp2} with the discrete pseudo-inverse operator.

\section{SPE and CSPE} 
\label{CSPE}
For the evaluation of the pseudo-inverse and for the application of the inverse of the matrix $\mathbf{M}_{cc}$ in \eqref{amp1} and \eqref{amp2} algebraic systems of equations are solved by the PCG method. Due to constant matrices $\mathbf{M}_{cc}$ and $\mathbf{K}_{nn}$ the corresponding systems of equations form multiple right-hand side (mrhs) problems.

Solutions for $\mathbf{k}_{p}$ from previous time steps are used to obtain an improved start vector for the PCG method. Within the subspace projection extrapolation (SPE) start vector generation method \cite{SPE04}, the solution
vectors from $m$ previous time steps are orthonormalized by the modified Gram-Schmidt process (MGS) to form the linearly independent column vectors of an operator $\mathbf{V}=\lbrace\mathbf{v}_{1}\mid...\mid\mathbf{v}_{m}\rbrace$. Solving the projected system:
\begin{equation}
\label{eqs_SPE}
\mathbf{V}^{\mathrm{T}}\mathbf{K}_{nn}\mathbf{V}\mathbf{z}=\mathbf{V}^{\mathrm{T}}\mathbf{r},
\end{equation} where $\mathbf{r}$ is the varying right-hand side vector, with a direct method yields the vector $\mathbf{z}\in\mathbb{R}^{m}$. This holds the coefficients for computing a new start vector
\begin{equation}
\label{x0_SPE}
\mathbf{x}_{0,\mathrm{SPE}}\mathrel{\mathop:} =\mathbf{V}\mathbf{z},
\end{equation} by linear combination of the earlier computed column vectors of the operator $\mathbf{V}$ \cite{SPE04}.

In this work, the SPE method is employed as follows: In the $(m+1)$-th time step, the solution from the last, i.e. $m$-th, time step is orthonormalized against the column vectors in $\mathbf{V}$ by MGS, and is referred to as $\mathbf{v^{m+1}}$. If the number of required PCG iterations is larger in the $m$-th time step than it has been in the $(m-1)$-th time step, $\mathbf{v^{m+1}}$ is appended as column vector to the operator $\mathbf{V}$. Otherwise, it is inserted into the last column of the operator $\mathbf{V}$. 
An increasing number of PCG iterations results from increasingly worse start vectors obtained with an operator $\mathbf{V}$ storing insufficient information with respect to the new vector $\mathbf{r}$. Appending $\mathbf{v^{m+1}}$ increases the spectral information content of $\mathbf{V}$ without deleting any information from previously included column vectors.
In the computation of the product $\mathbf{K}_{nn}\mathbf{V}$ in \eqref{eqs_SPE} only the matrix-column-vector product $\mathbf{K}_{nn}$ $\mathbf{v^{m+1}}$ changes in every time step. The other matrix-vector products $\mathbf{K}_{nn}$ $\mathbf{v_{i}}, i=1,...,m$, have been computed at previous time steps and can be reused. This modified version of the SPE is the so called "Cascaded SPE" (CSPE).

\section{Numerical Results}
\label{results}

\begin{figure}[h]
\includegraphics[width=\textwidth]{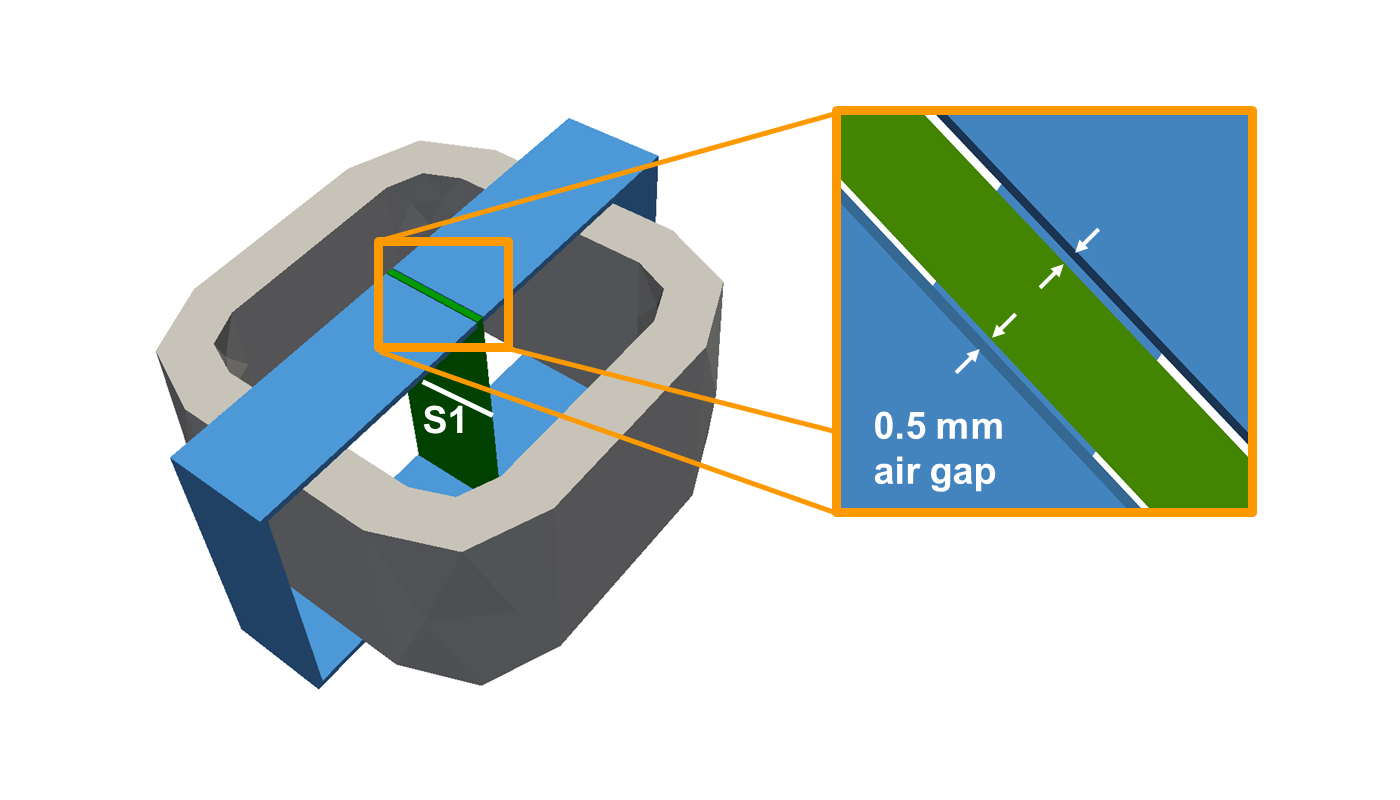}
  \caption{Geometry of the TEAM 10 model. The line S1 qualitatively shows where the average magnetic flux density is measured}
  \label{fig:team10}
\end{figure}

The nonlinear magnetoquasistatic TEAM benchmark problem "TEAM 10" \cite{Team10} is simulated. It consists of two steel plates in the shape of square brackets that are placed facing each other with a third, rectangular steel plate in their midst, with small air gaps in between, as depicted in  Figure \ref{fig:team10}.
At time $t=0\,s$, the current in the excitation coil starts to increase according to a $(1-\exp(-t/\tau))$ function and the reaction of the field of the magnetic flux density  was simulated for the first 120 ms.
The time step width for the explicit Euler method is bounded by
\begin{equation}
\label{maxDeltaT}
\Delta t\leq \dfrac{2}{\lambda_{\mathrm{max}}\left(\mathbf{M}_{cc}^{\mathrm{-1}}\left(\mathbf{K}_{cc}(\mathbf{a}_{c})-\mathbf{K}_{cn}\mathbf{K}^{\#}_{nn}\mathbf{K}^{\mathrm{T}}_{cn}\right)\right)},
\end{equation} as stated in \cite{Schur2}. For the maximum eigenvalue $\lambda_{\mathrm{max}}$ the proportionality:

\begin{equation}
\label{lambda}
\lambda_{\mathrm{max}}\left(\mathbf{M}_{cc}^{\mathrm{-1}}\left(\mathbf{K}_{cc}(\mathbf{a}_{c})-\mathbf{K}_{cn}\mathbf{K}^{\#}_{nn}\mathbf{K}^{\mathrm{T}}_{cn}\right)\right)\propto \dfrac{1}{h^{2}\kappa\mu},
\end{equation}holds, whereas $\mu$ is the permeability, $\kappa$ is the conductivity and $h$ is the smallest edge length of the mesh used for spatial discretization.

Eqn. \eqref{lambda} shows that a fine mesh resolution, e.g. of an narrow air gap, results in a small maximum stable CFL-time step criterion. A rather coarse mesh with 29,532 dofs is employed in order to allow for the variation of several parameters in an acceptable overall simulation time. The maximum stable time step width for this mesh is $\Delta t = 1.2\,\mu s$. As stated in Section \ref{CSPE}, the number of column vectors in the CSPE operator $\mathbf{V}$ is increased if the number of PCG iterations required in each time step is increasing. Preventing the number of column vectors in $\mathbf{V}$ from becoming too large, thus resulting in more expensive computations in the evaluation of \eqref{eqs_SPE}, is done by additionally setting a limit of accepted PCG iterations ($N_{CG,acc.}$). The number of column vectors in $\textbf{V}$ is only increased if $N_{CG,acc.}$ is exceeded. Parameters varied in the simulations are $N_{CG,acc.}$ and the tolerance for the stopping criterion for the PCG method when computing the pseudo-inverse. For each PCG tolerance $tol_{PCG}\in\lbrace 10^{-8}, 10^{-7}, 10^{-6}\rbrace$ simulations with $N_{CG,acc.}\in\lbrace 1, 3, 5\rbrace$ are run, in order to check the effect on the accuracy of the calculated magnetic flux density and on the simulation time.

\begin{figure}[htb]
\begin{minipage}[b]{0.5\textwidth}
\centering
  \includegraphics[width=\textwidth]{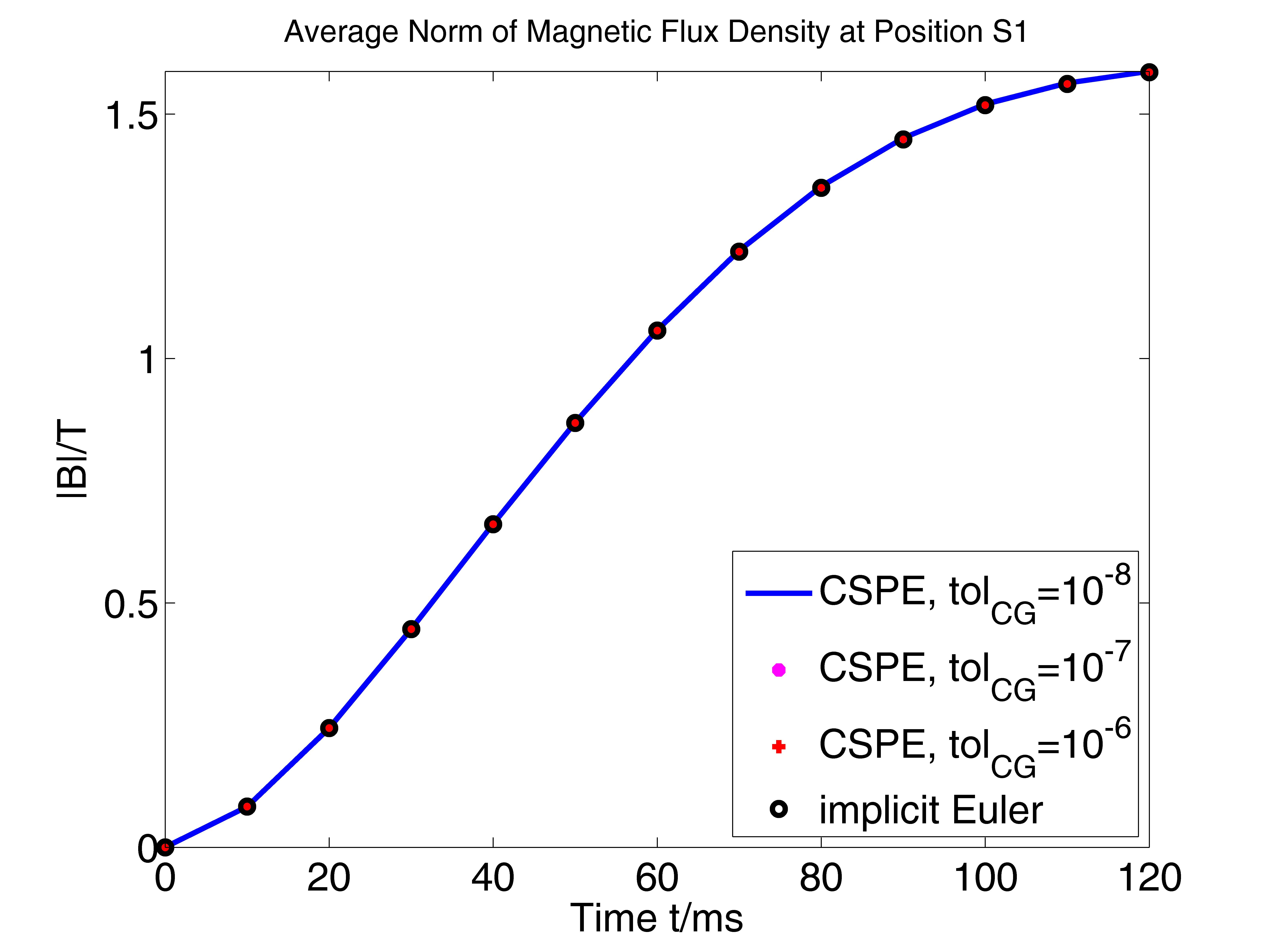}
  \caption{Comparison of results of implicit and explicit time integration at position S1}
  \label{fig:S1}
\end{minipage}
\hfill
\begin{minipage}[b]{0.5\textwidth}
\centering
  \includegraphics[width=\textwidth]{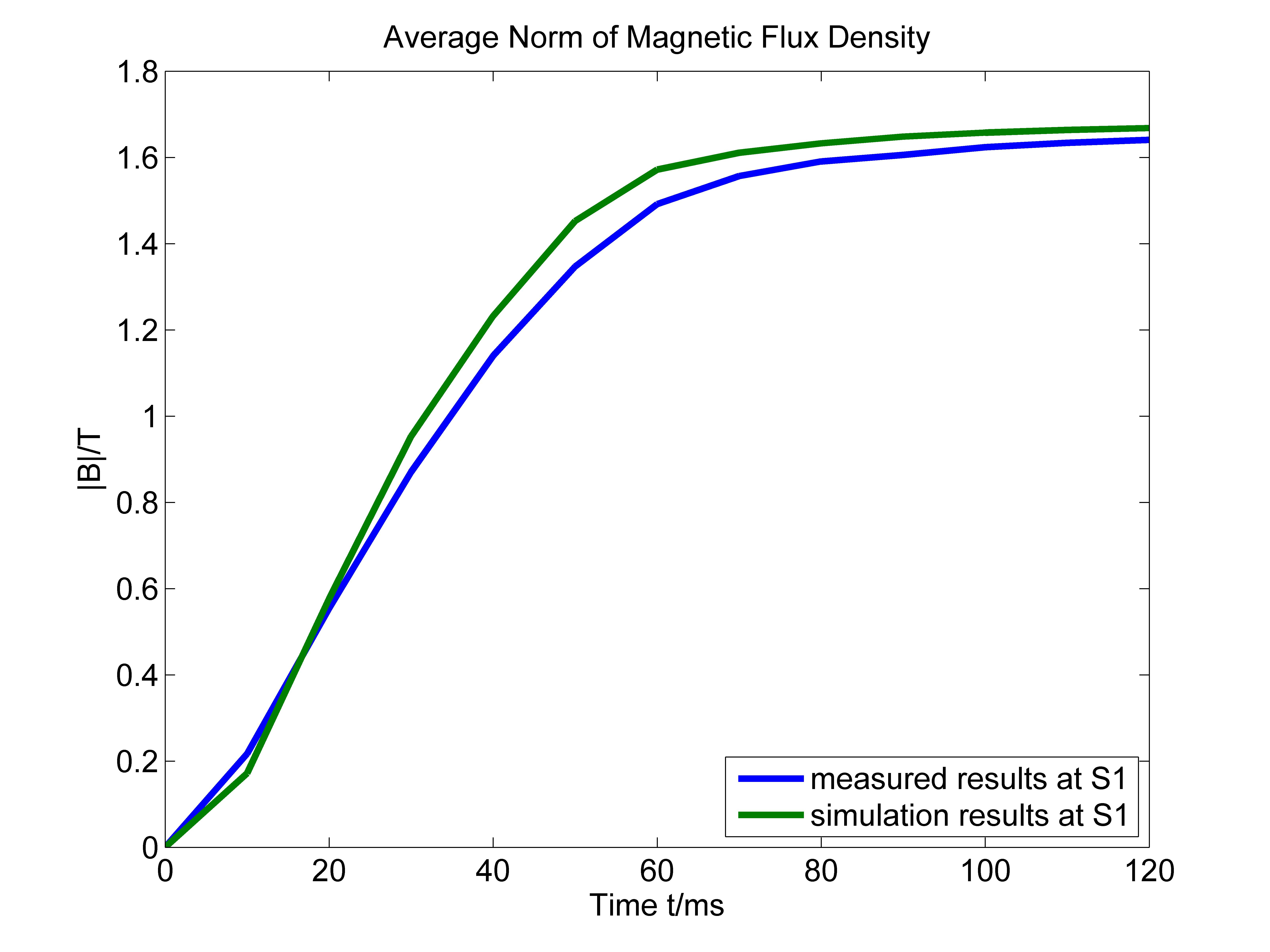}
  \caption{Comparison of simulation with about 700,000 dofs and measured results stated in \cite{Team10}}
  \label{fig:RefBFine}
\end{minipage}
\end{figure}

Such a coarse mesh does not yield sufficient accuray with respect to the measured reference results published in \cite{Team10}. Therefore, the results for the magnetic flux density obtained in simulations with the explicit time integration scheme are compared with the results of a reference simulation on the same mesh using the standard formulation and the implicit Euler method for time integration.
The results are depcited in Figure \ref{fig:S1} and show good agreement. In order to prove the accuracy of the employed code itself using an implicit time integration method, a finer discretization with about 700,000 dofs is chosen, resulting in a simulation time of 5.38 days on a workstation with an Intel Xeon E5-2660 processor. The results are presented in Figure \ref{fig:RefBFine}.
Independent of the time integration method, the spatial disretization is done by FEM based on $1^{\mathrm{st}}$ order edge elements.

\begin{figure}[htb]
\begin{minipage}[t]{0.5\textwidth}
  \includegraphics[width=\textwidth]{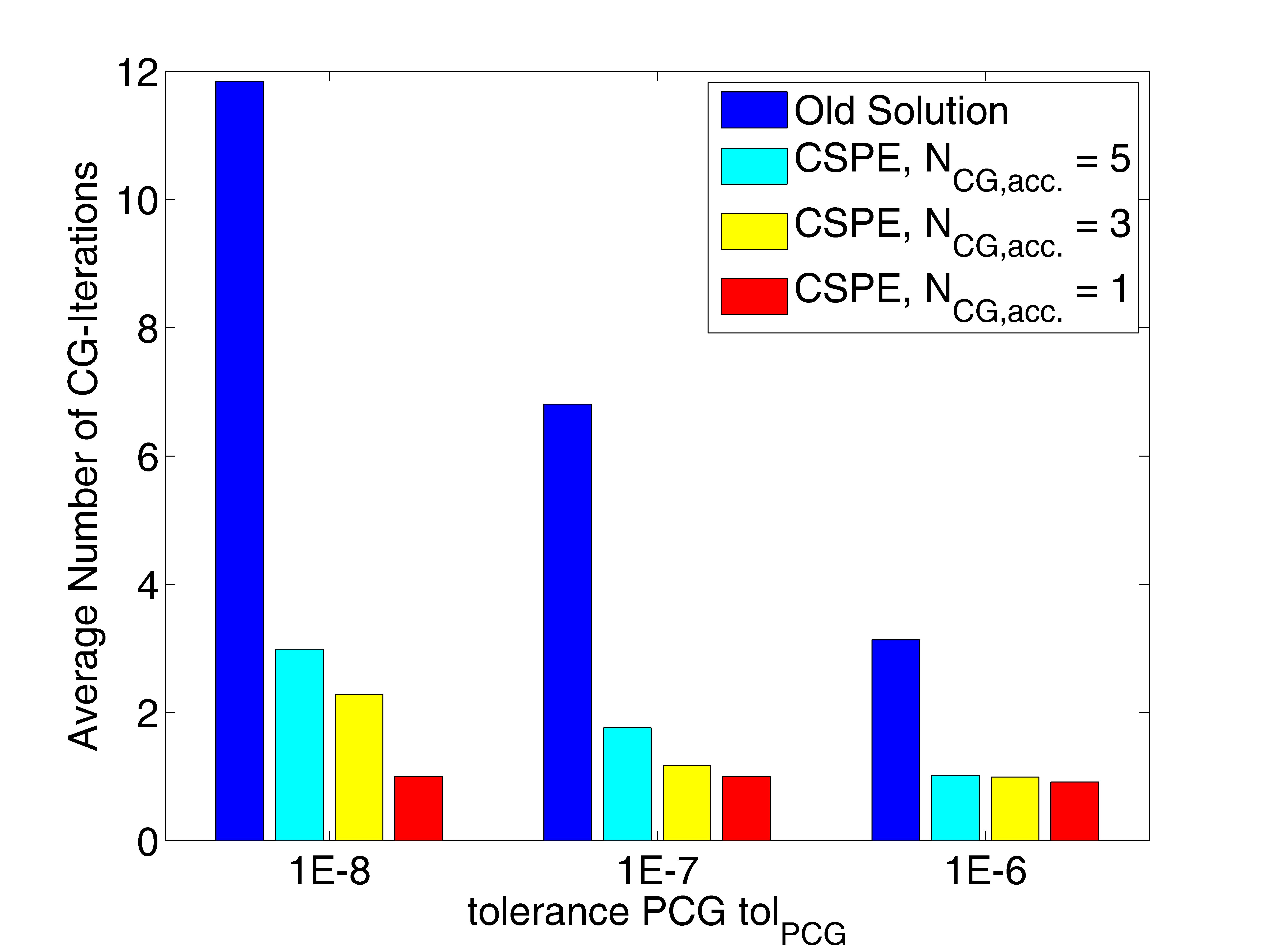}
  \caption{Average number of required PCG iterations}
  \label{fig:AvgCG}
\end{minipage}
\begin{minipage}[t]{0.5\textwidth}
  \includegraphics[width=\textwidth]{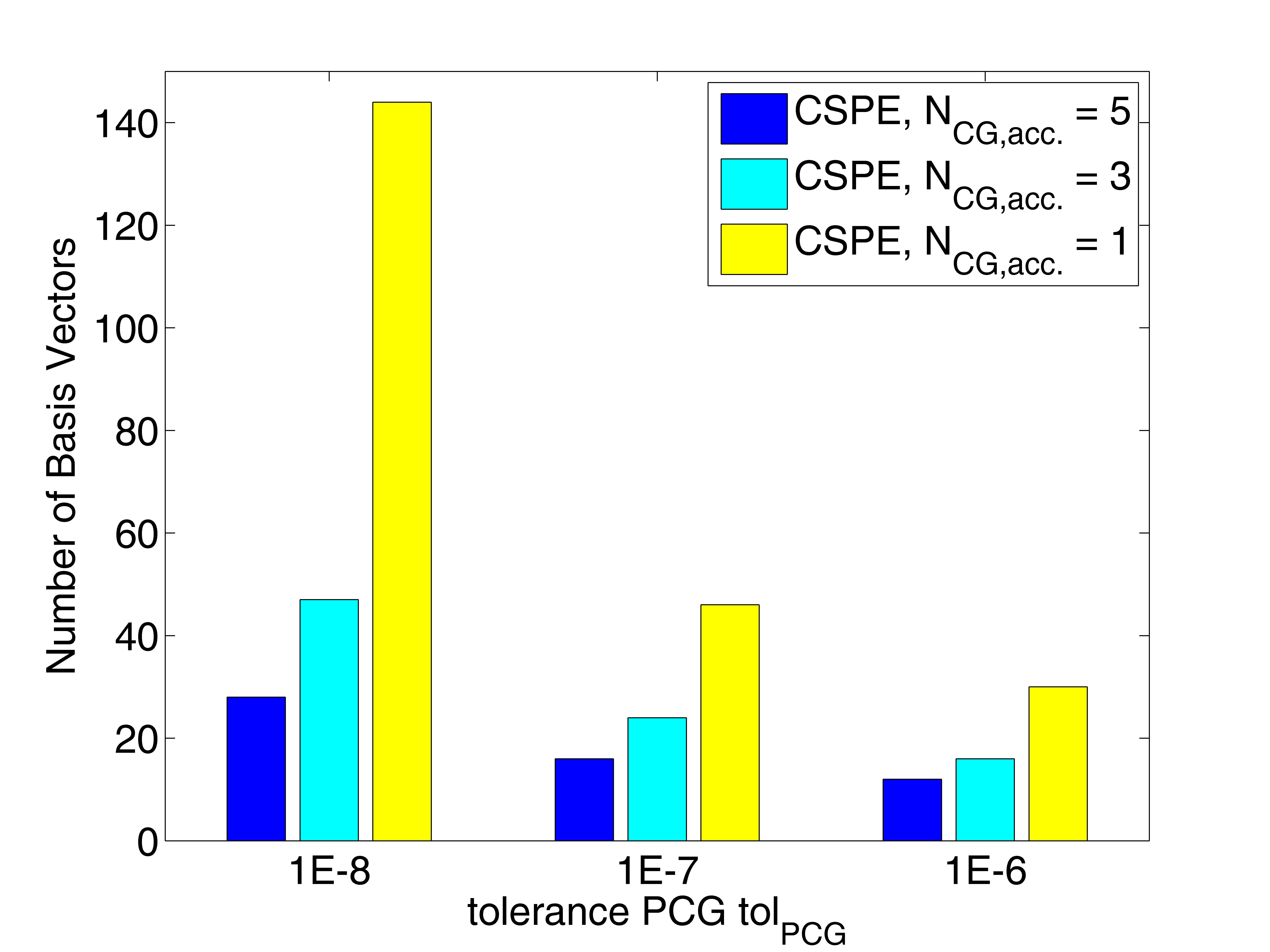}
  \caption{Maximum number of column vectors in the CSPE operator \textbf{V}}
  \label{fig:dimV}
\end{minipage}
\end{figure}

\begin{figure}[h]
\centering
\includegraphics[width=0.5\textwidth]{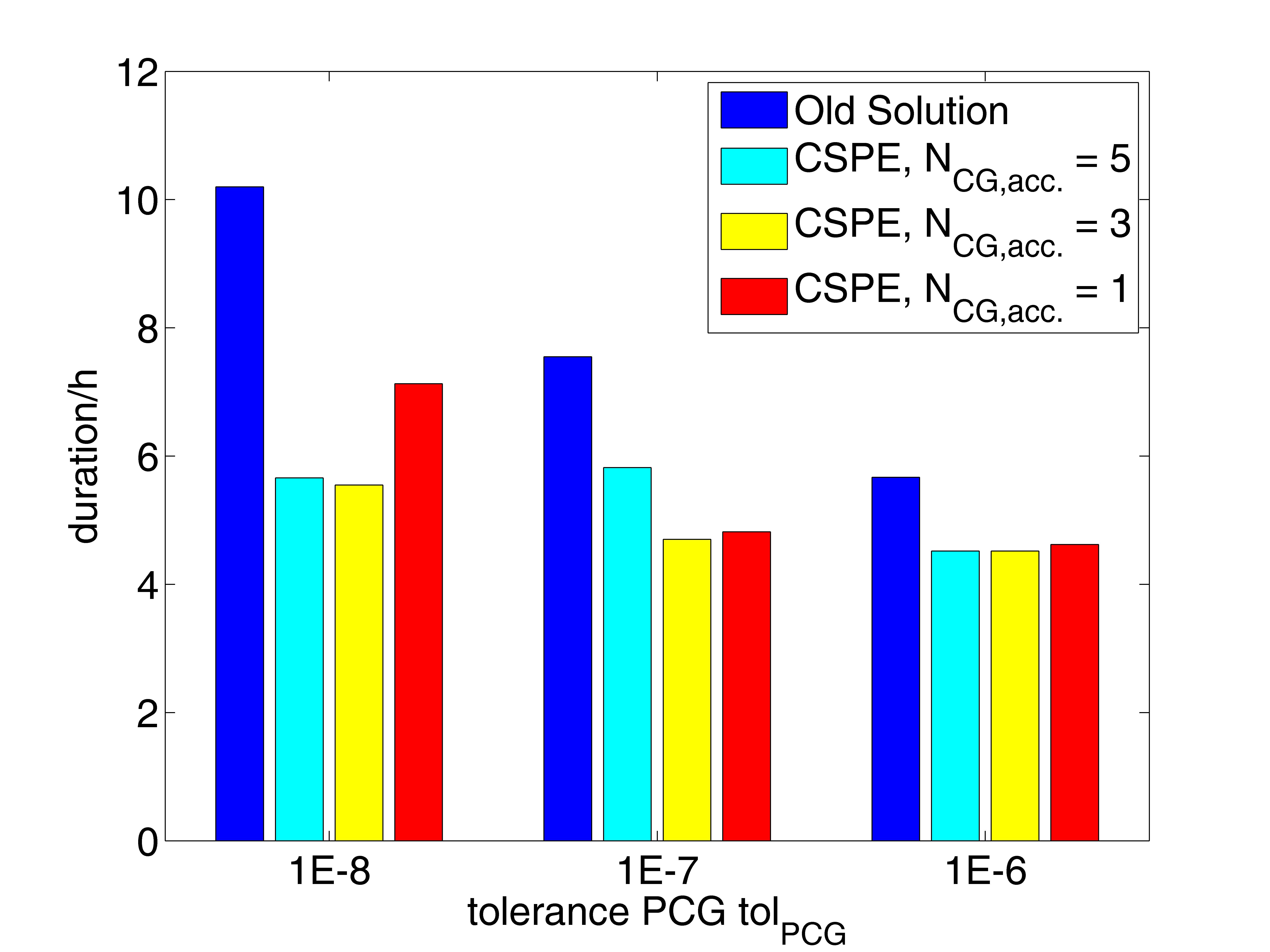}
  \caption{Duration of the simulations with explicit time integration}
  \label{fig:duration}
\end{figure}

The number of averagely used PCG iterations is decreased by reducing the prescribed PCG tolerance and by increasing the number of column vectors in the CSPE operator \textbf{V}, as becomes evident in Figures \ref{fig:AvgCG} and \ref{fig:dimV}. For each PCG tolerance, the use of the CSPE method reduces the average number of PCG iterations compared to using the solution from the previous time step as start vector by a factor of at least 4 up to a maximum of a factor 12. The effect of the number of column vectors in the operator \textbf{V} on the average number of required PCG iterations is less pronounced for lower PCG tolerances. Figure \ref{fig:S1}  demonstrates that there is no loss in accuracy by using a PCG tolerance of $10^{-6}$ when computing the pseudo-inverse for this test example. Figure \ref{fig:dimV} shows that a rather small number of column vectors of less than 20 in the CSPE operater \textbf{V} is sufficient. The simulation time is reduced by relaxing the PCG tolerance and using CSPE, as is shown in Figure \ref{fig:duration}. A decrease of simulation time with these two schemes by a factor of about 2.22 was achieved. However, the time needed for simulating this problem with an implicit Euler method was 2.35 h, so another speed-up of at least a factor of 1.9 will be necessary for the approach proposed in this method to become faster than the standard method using an implicit time integration scheme.

\section{Conclusion}
\label{conclusion}
A generalized Schur-complement was applied to the spatially discretized magnetic vector potential formulaiton of the eddy current problem. This transformed an inifitely stiff DAE(1) into an ODE of finite stiffness that was integrated in time using the explicit Euler method, thus avoiding linearization by the Newton-Raphson method.  Within the generalized Schur-complement approach  a pseudo-inverse for the singular curl-curl-matrix was evaluated by the PCG method. The average number of PCG iterations for evaluating this pseudo-inverse could be significantly reduced by generating an improved start vector for PCG with the CSPE method, which was demonstrated on simulations of the ferromagnetic TEAM 10 benchmark problem. Although the simulation time was reduced by use of the CSPE method, the main objective of being faster than the implicit time integration method should only become visible in case of easier to accomplish parallel implementation for massive many core systems. Further investigations should focus on further increasing the speed with which each time step is executed and on reducing the number of required time steps by using coarse space discretization combined with higher order schemes.

\section*{Acknowledgement}
This work was supported by DFG grants CL-143/11-1 and SCHO-1562/1- 1, the Excellence Initiative of German Federal and State Governments and the Graduate School CE at TU Darmstadt.

\end{document}